\documentclass[pra,aps,twocolumn,showpacs]{revtex4}
\usepackage{amsmath,amsfonts,amssymb,graphics,graphicx,epsfig,color,times}
\usepackage[latin1]{inputenc}
  
\newcommand{\bra}[1]{\left.\langle #1 \right|}
\newcommand{\ket}[1]{\left| #1 \right.\rangle}

\newcommand{\ad}{\hat{a}^\dagger}
\renewcommand{\a}{\hat{a}}
\newcommand{\nb}{\hat{n}}

\DeclareMathSymbol{\theta}{\mathord}{letters}{"23}
\DeclareMathSymbol{\rho}{\mathord}{letters}{"25}
\DeclareMathSymbol{\phi}{\mathord}{letters}{"27}
 
\DeclareMathSymbol{\vartheta}{\mathord}{letters}{"12}
\DeclareMathSymbol{\varphi}{\mathord}{letters}{"1E}
\DeclareMathSymbol{\varrho}{\mathord}{letters}{"1A}

\usepackage[usenames,dvipsnames]{xcolor}

\begin{document}
 
\newcommand{\nn}{{\mathbbm{N}}}
\newcommand{\rr}{{\mathbbm{R}}}
\newcommand{\cc}{{\mathbbm{C}}}
\newcommand{\id}{{\sf 1 \hspace{-0.3ex} \rule{0.1ex}{1.52ex}\rule[-.01ex]{0.3ex}{0.1ex}}}
\newcommand{\me}{\mathrm{e}}
\newcommand{\mi}{\mathrm{i}}
\newcommand{\md}{\mathrm{d}}
\renewcommand{\vec}[1]{\text{\boldmath$#1$}}

\newcommand{\Zak}{\text{Zak}}
\newcommand{\cmnt}[2]{{\color{blue} #1} {\color{red} \footnotesize \emph{\texttt{#2}}}}
\newcommand{\Cmnt}[1]{{\color{ForestGreen}#1}}
\newcommand{\cmntMH}[2]{{\color{RoyalBlue} #1}{\color{Red} \footnotesize \emph{\texttt{#2}}}}

\title{Topological edge states in the one-dimensional super-lattice Bose-Hubbard model}
\author{Fabian Grusdt$^{1,2}$\email{grusdt@physik.uni-kl.de}, Michael H\"oning$^{1}$\email{hoening@rhrk.uni-kl.de}, and Michael Fleischhauer$^{1}$}
\affiliation{$^{1}$ Dept. of Physics and Research Center OPTIMAS,  Technische Universit\"at Kaiserslautern, D-67663 Kaiserslautern, Germany \\
 $^{2}$ Graduate School Materials Science in Mainz,
  Technische Universit\"at Kaiserslautern, D-67663 Kaiserslautern, Germany} 
 
\begin{abstract} 
We analyze interacting ultra-cold bosonic atoms in a one-dimensional (1D)
super-lattice potential with alternating tunneling rates $t_1$ and $t_2$ and inversion symmetry, which is the bosonic analogue of the Su-Schrieffer-Heeger (SSH) model. 
A Z$_2$ topological
order parameter is introduced which is quantized for the Mott insulating
(MI) phases. Depending on the ratio $t_1/t_2$ the
$n=1/2$ MI phase is topologically non-trivial, which results in many-body edge states at open boundaries. In contrast to the SSH model
the bosonic counterpart lacks chiral symmetry and the edge states are no longer mid-gap.
This leads to a generalization of the bulk-edge correspondence, which we discuss in detail. The edge states can be observed in cold atom experiments by creating a step in the effective confining 
potential, e.g. by a second heavy atom species, which leads to an interface between two MI regions with filling $n=1$ and $n=1/2$. 
Shape and energy of the edge states as well as
conditions for their occupation are determined analytically in the strong coupling limit and 
in general by density-matrix renormalization group (DMRG) simulations.
\end{abstract}
 
 
\date{\today}
 
\maketitle
 
Topological phases have become an intensively studied
subject in many fields of physics. Key features of condensed-matter systems such 
as topological insulators \cite{Kane-PRL-2005,Bernevig-PRL-2006,top-insulator} or 
superconductors \cite{top-superconductors} as well as quantum Hall systems 
\cite{Klitzing-PRL-1980,Laughlin-PRL-1981,Thouless-PRL-1982,Moore-Read-NucPhys-1991}
have been related to robust edge states at interfaces between phases with different 
topological character \cite{Hatsugai1993}. This fundamental relation has also 
been found to hold true for various one-dimensional systems \cite{Kitaev2001,Ryu2002,Delplace2011,Kraus2012,Verbin2013}.
Ultra-cold atomic gases 
have developed into an ideal experimental testing 
ground for concepts of solid-state and many-body physics 
\cite{Dalibard-RMP} and they could become
important for studying topological effects
\cite{Alba-PRL-2011,Tarruel-Nature-2012,Goldman-PRL-2012,Atala2012}. One of the
simplest models possessing non-trivial topological properties is the inversion symmetric Su-Schrieffer-Heeger (SSH) model \cite{SSH1979}, 
which can be realized by ultra-cold 
fermions in a 1D tight-binding super-lattice (SL)
potential with alternating hopping amplitudes. 
Its topological properties are classified by a Z$_2$ topological invariant
given by the Zak phase \cite{Zak1989,Berry1984} and have been explored both theoretically \cite{Ryu2002,Delplace2011,Lang-PRL-2012} 
and recently experimentally \cite{Atala2012}.
Here we show that in the case of bosons, MI phases with filling $n=1/2$
can be non-trivial topological insulators as well, where the topological invariant is
the Z$_2$ many-body Berry phase, first introduced in this context by Hatsugai \cite{Hatsugai2006}.
It has also been pointed out in \cite{Hatsugai2006,Hatsugai2007} that the Haldane phase \cite{Haldane1983} 
can be characterized by a similar Z$_2$ Berry phase. 
This system is well known to support topological many-body edge states \cite{Kennedy1990}, which we 
take as motivation to study the relation between the quantized Berry phase and topological edge states of the SL-Bose-Hubbard model (SL-BHM).

For the case of an ultra-cold bosonic lattice gas, introducing a
localized potential step allows to create an interface between gapped
MI phases with different topological invariants.
Due to the interface, many-body ground states emerge that display density minima or maxima
at the interface in analogy to an unoccupied or occupied single-particle edge state for free fermions. This 
can easily be observed with techniques developed in recent years \cite{Gericke-NatPhys-2008,Bakr-Nature-2009,Sherson-Nature-2010}. 
While for the SSH model a strict relation between the existence of a single-particle mid-gap edge state
at open boundaries and the bulk topological invariant has been identified \cite{Ryu2002,Delplace2011}, a
similar relation does in general not hold for the bosonic SL model with finite interactions
due to the absence of chiral symmetry. Instead, as 
we will show using numerical DMRG simulations \cite{White1992,Verstraete2006,Schollwock2011} and analytic 
approximations, a generalized bulk-edge correspondence 
holds: While either the empty (hole) or the occupied (particle) edge state remain localized and thus stable until the MI melts due to tunneling, one of the two many-body states hybridizes with the bulk already for much smaller
values of the tunneling rate.

\begin{figure}[hbt]
 \centering
\epsfig{file=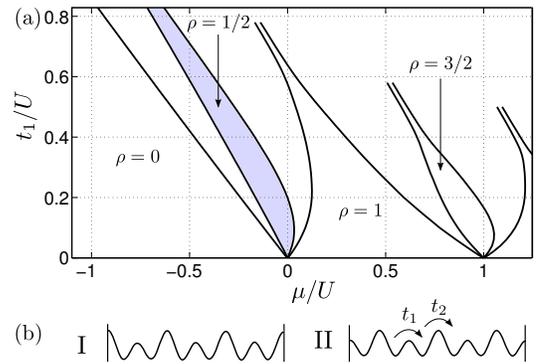,width=0.8\columnwidth}
 \caption{(Color online) (a) Phase-diagram for the SL-BHM with 
$t_2 = 0.2 t_1$ obtained by DMRG. One recognizes the presence of MI phases with
integer and half-integer filling.
(b) Different dimerizations I and II of SL potential corresponding to
Berry phases $\nu^\text{I}=0$ and $\nu^{\text{II}}=\pi$.} 
 \label{fig:phase_diagram_hopping}
\end{figure}

The starting point of the discussion is the 1D SL-BHM
in the grand-canonical ensemble described by the 
Hamiltonian $\hat K=\hat H-\mu \hat N$
\begin{eqnarray}
\hat{K}&=&-\sum_{j \text{ odd}} \left(t_1\ \ad_{j}\a_{j+1}+\rm{h.a.}\right)-\sum_{j \text{ even}} \left(t_2\ \ad_{j}\a_{j+1}+\rm{h.a.}\right)\nonumber\\
&& +\frac{U}{2}\sum_j\nb_j\left(\nb_j-1\right) +\sum_j (\epsilon_j-\mu) \nb_j,\label{eq:hamiltonian}
\end{eqnarray}
where $\a_j$ and $\ad_j$ are the annihilation and creation operators at lattice site
$j$, and $\nb_j=\ad_j\a_j$. Particles can tunnel with alternating hopping amplitudes $t_1$ and $t_2$ and 
there is an on-site interaction  $U$. $\epsilon_j$ describes a potential and $\mu$ is the chemical potential. 
In the case of hard-core bosons $(U\to \infty$) and $\epsilon_j\equiv \text{const}$, Eq. (\ref{eq:hamiltonian}) 
is equivalent to the inversion symmetric SSH model.

A generic ground-state phase diagram of the SL-BHM, taken from Ref.\cite{Muth-PRA-2008},
 is shown in Fig.\ref{fig:phase_diagram_hopping} for $\epsilon_j\equiv 0$. Besides MI phases with integer filling it shows loophole insulating regions with half integer filling for $t_1 > t_2$
\cite{Buonsante-PRA-2004,Buonsante-PRA-2005}. These 
regions shrink when $t_1$ decreases, and vanish at $t_1=t_2$ (simple BHM). They reappear when $t_1 < t_2$ and the point $t_1=t_2$ marks a topological phase transition. The model possesses chiral symmetry at half-integer filling in the limit $U\to\infty$.

In the case of non-interacting fermions, the topology of the band structure is
determined by its Zak phase \cite{Zak1989} (or winding number),
$ \nu = i \int_0^{2 \pi} dk \bra{u(k)} \partial_k \ket{u(k)}$, where
$|u(k)\rangle$ are the single-particle Bloch functions and $k$ is the lattice quasi-momentum.
While for a general 1D band structure this phase may take arbitrary values, it is 
integer (i.e. Z) quantized (in units of $\pi$) \cite{Ryu2010} when 
chiral symmetry is present, and Z$_2$ quantized for the SSH case.
In this case the winding numbers of the upper and lower band are equal but opposite  $\nu=0$ ($\pm \pi$) 
for dimerization I (II), see Fig. \ref{fig:phase_diagram_hopping}.

For interacting systems there is no conserved lattice quasi-momentum $k$ and one must employ a many-body generalization of 
the winding number. Like the Chern number \cite{Niu1985} it can be defined via generalized boundary conditions \cite{Resta1995},
$\psi(x_j+L)=e^{i \vartheta} \psi(x_j)$ for all
coordinates $j=1,...,N$ and system size $L$. These
correspond to a magnetic flux $\vartheta$ threading the system. When this flux is adiabatically varied, the many-body wavefunction 
$\ket{\Psi(\vartheta)}$ picks up a Berry phase \cite{Berry1984}
\begin{equation}
 \nu=i\int_0^{2\pi}\!\!\ d\vartheta \bra{\psi(\vartheta)} \partial_\vartheta \ket{\psi(\vartheta)}.
 \label{eq:ZakPhase}
\end{equation}
This topological order parameter can easily be calculated in the hard-core limit $U \to \infty$. 
We find that the MIs with integer filling are topologically trivial with $\nu=0$ and those with 
half-integer filling can take the values $\nu^{\rm I}=0$ for dimerization I and $\nu^{\rm II}=\pi$
for dimerization II. This is a direct consequence of the quantization of the free-fermion winding 
number for Hamiltonians possessing chiral symmetry \cite{Ryu2002,Ryu2010}. Most importantly the 
topological invariant stays strictly quantized even for finite $U$ as long as the particle-hole gap 
is finite. This however is a consequence of inversion symmetry alone and was realized already by Zak 
\cite{Zak1989}. An exact proof including the interacting case can be given 
following the proof of Hatsugai \cite{Hatsugai2006} \footnote{Our system is invariant under both 
spatial inversion ($P$) and time-reversal ($T$) even when twisted boundary conditions are used. Then the
combined symmetry $P T$ is of the type considered in \cite{Hatsugai2006}.}
. We checked the quantization for small systems by exact diagonalization, but the Z$_2$ 
invariant could as well be calculated using DMRG or as was recently shown using quantum Monte Carlo 
\cite{Motoyama2013}. Thus we expect the non-trivial topology of the SSH bands 
to carry over to bosons with finite interactions. This is our motivation to study 
edge states of topologically non-trivial
MI phases in the SL-BHM as indicators for a quantized topological invariant.

\begin{figure}[hbt]
 \centering
\epsfig{file=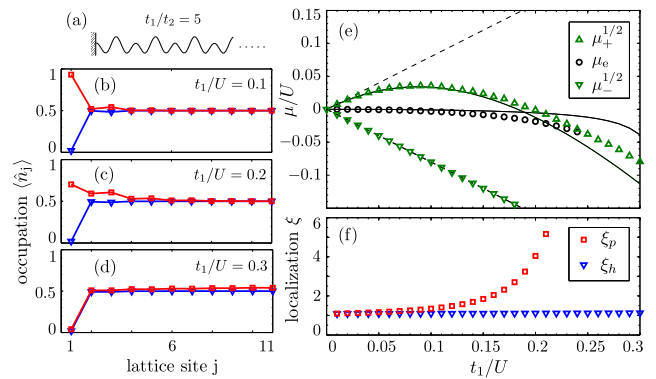,width=\columnwidth}
 \caption{(Color online) (a) $n=1/2$ MI with open boundary corresponding to
topologically non-trivial dimerization (II). Plots in (b)-(d), calculated by DMRG, show 
density distributions below (blue triangles) and above (red squares) the critical chemical 
potential of edge-state occupation $\mu_e$ for $t_1/U=0.1, 0.2, 0.3$ respectively. 
While a well localized hole state (empty edge) can be observed in all cases, the 
particle state (occupied edge) becomes unstable already before the MI melts (e). Due to the absence of chiral
symmetry $\mu_+^{1/2}$ (green up-pointing triangles) approaches $\mu_e$ (black circles) already at small
values of $t_1/U$.  Solid curves show analytic results from CSCPE, dashed straight lines correspond to hard-core
results. (f) When $\mu_e$ approaches $\mu_+^{1/2}$ the particle state becomes delocalized as can be see from the 
localization length $\xi_{p,h}$ of particle and hole edge states. Systems of length $L=65$ are considered in the 
DMRG simulation and numeric error bars are within the symbol size.
} 
 \label{fig:edge-states}
\end{figure}

Let us consider a $n=1/2$ MI. For a chemical potential within the range
\begin{equation}
 \mu_-^{1/2} < \mu < \mu_+^{1/2}
\end{equation}
the bulk is in a gapped phase with half filling. 
For large interactions $U\gg t_1>t_2$ the values of $\mu_\pm^{1/2}$ can be determined perturbatively
within a ``cell strong-coupling perturbative expansion'' CSCPE \cite{Freericks1996,BPV-PRB-2004}
up to order ${\cal O}(t_2^2/U,\, t_2t_1^2/U^2)$:
\begin{eqnarray}
 \mu_-^{1/2} &=& - (t_1-t_2),\\ 
\mu_+^{1/2} &=& (t_1-t_2 ) +\frac{U}{2}-\frac{1}{2}\sqrt{16t_1^2+U^2} -\frac{4 t_1t_2}{U}.
\end{eqnarray}
Choosing e.g. a ratio $t_1/t_2=5$ the bulk gap 
$\Delta=\mu_+^{1/2} - \mu_-^{1/2}$ remains finite until about $t_1/U \approx 1.2$.
In the limit of large $U$ there is chiral symmetry and $\mu_-^{1/2} = - \mu_+^{1/2}$.
In this limit
we expect from analogy with the SSH model a mid-gap edge state if we add an open boundary with the
topologically non-trivial dimerization (II),  see Fig. \ref{fig:edge-states}(a). 
And indeed for small values of $t_1/U=0.1$ and $0.2$ DMRG simulations show both a well 
localized hole ($\psi_h$) and particle ($\psi_p$) state below and above a critical chemical 
potential $\mu_e$, see Fig.\ref{fig:edge-states}(b) and (c). These grand canonical ground states differ 
in their total particle number by one. However, as shown in Fig.\ref{fig:edge-states}(d), already for 
$t_1/U =0.3$, i.e. well before the MI melts due to tunneling the situation changes: when increasing the chemical potential
the density of the bulk increases non-locally instead of filling up the hole at the edge.
Interestingly the localized hole at the edge survives even when the chemical potential exceeds 
$\mu_+^{1/2}$ and the bulk becomes gapless. We will not discuss this here.
Within  CSCPE we find for the critical chemical potential $\mu_e$ where the grand canonical
ground state turns from $\psi_h$ to $\psi_p$
\begin{equation}
 \mu_e = - 2t_2^2 \frac{U-2t_1}{(U+t_1)(U-3 t_1)}
\end{equation}
We have plotted this result for $\mu_e$ along with the values from DMRG simulations 
in Fig.\ref{fig:edge-states}(e). As chiral symmetry is broken for finite values of the 
interaction, $\mu_e$ is no longer exactly in between $\mu_-^{1/2}$ and $\mu_+^{1/2}$. Furthermore
at a tunneling rate of $t_1/U \approx 0.25$ the curve touches $\mu_+^{1/2}$ indicating that it becomes 
energetically favorable to add a particle to the bulk rather than to the empty edge state (hole state).  
Within CSCPE we find  for the critical value
\begin{equation}
 \left(\frac{t_1}{U}\right)_c \approx \frac{1-\eta}{4(1+\eta -\eta^2/2)},\quad \eta=\frac{t_2}{t_1},
\end{equation}
\noindent which is slightly below the numerical value.

One recognizes that the curve of $\mu_e$ remains almost a straight line and starts to bend only
when it approaches $\mu^{1/2}_+$. This is due to an increasing delocalization of the 
particle edge state $\psi_p$. In Fig.\ref{fig:edge-states}(f) we have plotted
the numerically determined localization length $\xi_{p,h}$ for the particle and hole 
states in units of the lattice constant,
defined through the participation ratio $\xi=\bigl(\sum_j \Delta n_j\bigr)^2/\sum_j \bigl(\Delta n_j \bigr)^2$ where 
$\Delta n_j =|n_j-\frac{1}{2}| \Theta \left( \pm ( n_j - \frac{1}{2} ) \right)$
with ``$+$'' for $\xi_p$ and ``$-$'' for $\xi_h$ and where $\Theta$ is the Heaviside step function \cite{Murphy2011}.
While the hole state remains well localized, the localization
length of the particle state diverges as the tunneling rate approaches the critical value $(t_1/U)_c$.

Although the bulk-edge correspondence does not hold in the sense of a 
protected and localized many-body \emph{mid-gap state}, we found that the edge features 
of topologically trivial and non-trivial phases at an open boundary are markedly 
different. 
Although, in the topologically non-trivial case, the localized particle feature disappears at $(t_1/U)_c$ we 
generally find that \emph{at least one} of the particle and hole states remains stable. This holds true for all 
parameters corresponding to a gapped MI in the bulk.
This is a direct consequence of topology: A non-trivial bulk of Berry phase 
$\nu=\pi$ can be reached from the trivial bulk with $\nu=0$ only through a 
topological phase transition. However when the underlying symmetries (inversion in our case) are broken 
one phase can be adiabatically transformed into another, which corresponds to a \emph{quantized, half} 
Thouless pump (TP) cycle \cite{Thouless-PRB-1983}. Since the Berry phase changes by $\Delta \nu=\pi$, 
the polarization, i.e. the center of mass must change by one lattice site. This follows 
from the one-to-one relation between Berry phase eq.\eqref{eq:ZakPhase} and polarization \cite{Kingsmith1993,Ortiz1994}.
This argument is strict in an infinite system. It carries over to semi-infinite systems with a single 
open boundary, say on the left (negative) side, provided the systems many-body gap still remains finite during the 
entire TP cycle. In this case the pump effectively creates a hole of charge $-1/2$ 
localized on the open boundary. (I.e. the density relative to the topologically trivial case decreases by 
an amount corresponding to half a particle). The same holds true for a particle of charge $1/2$ 
when the TP is reversed and $\Delta \nu = -\pi$. Since the relevant gaps 
in the particle (hole) case are $\mu^{1/2}_+-\mu_e$ ($\mu_e-\mu^{1/2}_-$), both are stable for $t_1/U < (t_1/U)_c$.
Furthermore at least one must be stable as long as the bulk particle-hole gap $\mu_+^{1/2}-\mu_-^{1/2}$ is positive.

\begin{figure}[t]
 \centering
\includegraphics[width=0.90\columnwidth]{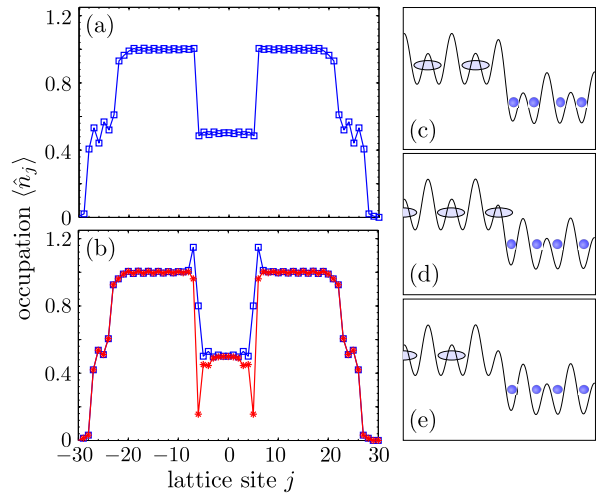}
 \caption{(Color online)  Ground-state density distribution of
the SL-BHM with harmonic trap and potential step 
between sites $j= - 6$ and $j= 5$ leading to interfaces between $n=1/2$  (in the center)
and $n=1$ MI regions. $\epsilon_j^\text{trap}=\omega (j+0.5)^2$, with $\omega/U=0.001$. Results are obtained by DMRG simulations for
$\mu/U = 0.55$.
(a) topological trivial $n=1/2$ MI phase with $t_1/U = 0.04$, $t_2/U = 0.2$ and $\Delta\epsilon/U =0.6$.
(b) topological non trivial $n=1/2$ phase  with $t_1/U = 0.2$, $t_2/U = 0.04$ and $\Delta\epsilon/U =0.6 (\text{blue squares}), 0.7 (\text{red stars})$. 
Right panel illustrates interface in topologically trivial case (c) and in nontrivial case with  
occupied (d) and unoccupied interface (e).}
 \label{fig:DMRG-trap}
\end{figure}

In the following we discuss a possible experimental realization of edge states 
using ultra-cold atoms. Although a sharp open boundary is difficult to realize, an interface between 
two MI phases with integer (e.g. $n=1$) and half-integer filling (e.g. $n=1/2$)
can be created by increasing the potential energy $\epsilon_j$ by $\Delta\epsilon$ for a number of 
consecutive lattice sites such that 
\begin{eqnarray}
  \mu_-^1 \ <  & \mu & <   \mu_+^1 \label{eq:cond1} \\
 \mu_-^{1/2} + \Delta \epsilon  < & \mu & < \mu_+^{1/2} + \Delta \epsilon.\label{eq:cond2}
\end{eqnarray}
Here $\mu^{1/2}_\pm$ and
$\mu^1_\pm$ denote the upper $(+)$ and lower $(-)$ boundaries of the insulating regions
in the phase diagram of Fig.\ref{fig:phase_diagram_hopping}.
As shown e.g. in Ref.\cite{Mering-PRA-2008} for the case of Bose-Fermi mixtures, an effective  potential step 
can be created by an admixture of a second atomic species, e.g. fermions, with very small hopping rates.
Under appropriate conditions (see \cite{Mering-PRA-2008}) 
the fermions form a connected cluster at the center of the trap with unity filling and sharp boundaries.
This results in an increase of the potential energy of the bosons $\Delta \epsilon$ which 
extends over all sites of the fermion cluster.
Depending on the location of the interfaces 
relative to the sub-lattices the winding number either stays,
$\Delta \nu =\nu_{1/2}^\text{I}-\nu_1 =0$, or jumps, $\Delta \nu =\nu_{1/2}^{\text{II}}-\nu_1 = \pi$.

Fig.\ref{fig:DMRG-trap} shows the density distribution in a weak harmonic trap 
with an additional potential step $\Delta\epsilon$ calculated by DMRG.
One clearly recognizes interfaces between a central $n=1/2$ MI and surrounding $n=1$ 
MI regions. Since the number of heavy particles was taken to be even, both interfaces 
are characterized by the same change $|\Delta\nu|$. The upper plot shows the case
$\Delta\nu =0$, the lower one $\Delta\nu=\pi$.  In the first case there is a simple step in the
density and no additional structure at the edge. The same holds at an interface between any two MI phases
with integer fillings irrespective of the dimerization.
In the second case, however, one sees pronounced
dips or peaks in the average density.

Generalizing the CSCPE to the case of an interface with
a finite potential step one can easily determine the potential 
heights $\Delta \epsilon$ for which non of the states $\psi_{h,p}$ hybridizes 
with the bulk, as well as the critical chemical potential at which the 
many-body ground state turns from the hole edge state $\psi_h$ to the particle 
edge state $\psi_p$. The first case ($\psi_h$) can be detected by measurement of a local particle number less than $1/2$ 
on the $n=1/2$ MI side of the interface, the second ($\psi_p$) by measurement of a local particle number larger than $1$ on the $n=1$ MI side.

\begin{figure}[t]
 \centering
\includegraphics[width=0.95\columnwidth]{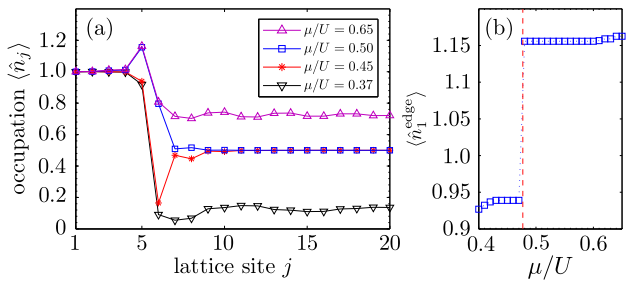}
 \caption{(Color online)  (a) Density distribution at potential step $\Delta\epsilon/U=0.6$ at $j_\text{step}=6$ for $t_1/U=0.2$, $t_1/t_2=5$ 
and increasing chemical potential $\mu/U$. (b) Local particle number $\langle \hat{n}_1^\text{edge}\rangle$ at the edge 
of the $n=1$ MI region as function of $\mu/U$ showing the occupation at the edge 
if $\mu > \mu_e=0.47$ (dashed red line).} 
 \label{fig:DMRG-edge}
\end{figure}

To verify these results we performed DMRG simulations for a step potential $\epsilon_j/U=\Delta\epsilon 
\Theta(j-j_\text{step}+0.5)$. In Fig.\ref{fig:DMRG-edge}a we show the density for different values 
of $\mu$. For $\mu/U=0.45$ the system is inside the stability region of both edge states. 
One clearly recognizes a well localized dip in the density. $\mu/U=0.50$ corresponds 
to an occupied edge inside the stability region. Here a clearly pronounced density peak appears. When $\mu$ 
is chosen such that the system is outside the region of the $n=1/2$ Mott insulator 
($\mu/U = 0.37$ and $\mu/U=0.65$) the density dip on the $n=1/2$ side starts to vanish while interestingly
the peak on the $n=1$ MI side remains. Fig.\ref{fig:DMRG-edge}(b) shows the local occupation number 
$\langle \hat{n}_1^\text{edge}\rangle$ at the edge of the $n=1$ MI side as function of $\mu/U$. As soon as 
$\mu$ exceeds $\mu_e$ as calculated in CSCPE (red dashed line), there is a clear jump indicating the transition 
from hole to particle edge state.

A particular feature of the edge state in the SL-BHM, not present in the hard-core limit,
is the peak of the local density on the border of the $n=1$ MI region above unity. Within CSCPE 
we calculate this density, which yields in zeroth order of $t_2$
\begin{equation}
 \langle \hat{n}_{1}^{\rm edge} \rangle= 1+\frac{4 \Delta \epsilon 
 t_1^2}{U^3}+\frac{6 \Delta \epsilon ^2
   t_1^2}{U^4}+\mathcal{O}(1/U^5).
\label{eq:edge-density}
\end{equation}
We checked the validity of this result by comparing to DMRG data and found good agreement until 
the $n=1$ MI starts to melt.

In summary we have discussed topological properties of the one dimensional SL-BHM
with alternating hopping rates $t_1$ and $t_2$. In the limit of infinite interaction $U$ this model corresponds to the SSH model for free fermions, which is known to possess topologically
non-trivial insulating phases for $t_1\ne t_2$ with a strict bulk-edge correspondence. We 
introduced a many-body generalization of the Zak phase
as topological order parameter, which is quantized as a consequence of inversion symmetry.
We analyzed edge states of a MI with filling $n=1/2$ 
for open boundary conditions using DMRG and analytic perturbative
calculations and found that the bulk-edge correspondence does not hold strictly in the sense of
a protected and localized many-body mid-gap state. Instead we showed that, as a direct consequence
of non-trivial topology, at least a particle- or a hole-like edge state remains localized and stable 
until the MI melts. While sharp open boundaries may be difficult to realize in cold-atom experiments, we showed that an interface between a $n=1$ and $n=1/2$ MI can be created where two topologically distinct phases are in contact. The required potential step can be realized by an admixture
of a second heavy atom species. We found that similar edge states emerge as in the case of
open boundary conditions. These edge states are characterized by a density dip at the edge 
below $1/2$ and a density peak at the edge 
with local particle number exceeding $1$. These features allow a simple detection
of the edge states and thus a verification of the different 
topological nature of the MI phases in cold-atom experiments.

The authors thank E. Demler and J.Otterbach for stimulating discussions and the referee for valuable input.
F.G. thanks E. Demler and the physics department of Harvard University for hospitality 
during his visit and the graduate school MAINZ for financial support. Financial support from 
the research center OPTIMAS is gratefully
acknowledged.

\end{document}